\documentclass[10pt,conference]{IEEEtran}
\usepackage{amsmath,amsfonts}
\usepackage{algorithmic}
\usepackage{algorithm}
\usepackage{array}
\usepackage[caption=false,font=normalsize,labelfont=sf,textfont=sf]{subfig}
\usepackage{textcomp}
\usepackage{stfloats}
\usepackage{url}
\usepackage{verbatim}
\usepackage{graphicx}
\usepackage{cite}
\usepackage[numbers,sort&compress]{natbib}
\usepackage{booktabs}
\usepackage{graphics}
\usepackage{adjustbox}
\usepackage{colortbl}
\usepackage{subcaption}
\usepackage{xspace}
\usepackage{listings}
\usepackage{multirow}
\usepackage{enumitem}
\usepackage{xcolor}
\usepackage{hyperref}
\usepackage[normalem]{ulem}
\usepackage{wrapfig}
\usepackage{amsmath}
\usepackage[most]{tcolorbox} %

\usepackage{tikz}
\usetikzlibrary{shapes.misc, positioning}
\usepackage{fontawesome5}

\definecolor{comment}{rgb}{0,0.6,0}
\definecolor{dgray}{gray}{0.25}
\definecolor{strings}{RGB}{0,0,128}
\definecolor{backcolour}{rgb}{0.95,0.95,0.92}
\definecolor{keywords}{RGB}{127,0,85}
\definecolor{darkred}{RGB}{139,0,0}
\definecolor{darkyellow}{RGB}{204,204,0}
\definecolor{darkblue}{rgb}{0.0, 0.0, 0.55}
\definecolor{vividviolet}{rgb}{0.62, 0.0, 1.0}
\definecolor{fuchsia}{rgb}{1.0, 0.0, 1.0}
\definecolor{shockingpink}{rgb}{0.99, 0.06, 0.75}
\definecolor{darkBlue}{RGB}{0, 0, 205}
\definecolor{gray85}{gray}{0.85}
\definecolor{nightBlue}{RGB}{45,78,159}

\definecolor{nordGreen2}{RGB}{152, 187, 186}
\definecolor{nordGreen}{RGB}{163, 190, 140}
\definecolor{nordPurple}{RGB}{180, 142, 173}
\definecolor{nordOrange}{RGB}{208, 135,112}

\definecolor{newred}{rgb}{0.0, 0.0, 1.0}  %
\definecolor{newgreen}{rgb}{0.5, 0.0, 0.5}  %
\usepackage{hyperref}
\hypersetup{
    colorlinks=true,
    linkcolor=nightBlue,       %
    citecolor=nightBlue,     %
    urlcolor=nightBlue       %
}

\newcommand{\mycomment}[1]{}

\newcommand{\dm}{developer-created module\xspace}
\newcommand{\dms}{developer-created modules\xspace}

\newcommand{\da}{developer-created architecture\xspace}
\newcommand{\das}{developer-created architectures\xspace}

\newcommand{\rem}{recovered module\xspace}
\newcommand{\rems}{recovered modules\xspace}
\newcommand{\rea}{recovered architecture\xspace}
\newcommand{\reas}{recovered architectures\xspace}

\newcommand{\jms}{{Java modules}\xspace}

\newcommand{\jpms}{JPMS\xspace}
\newcommand{\llm}{\textit{ClassLAR}\xspace}
\newcommand{\llmfull}{\textbf{Class}- and \textbf{L}anguage model-based \textbf{A}rchitectural \textbf{R}ecovery\xspace}

\newcommand{\hscore}{\textit{h-score}\xspace}
\newcommand{\hscores}{\textit{h-scores}\xspace}
\newcommand{\cscore}{\textit{c-score}\xspace}
\newcommand{\cscores}{\textit{c-scores}\xspace}
\newcommand{\ctoc}{\textit{c2c}\xspace}

\definecolor{findingbg}{RGB}{234,240,252}
\newcounter{findingcounter}
\newtcolorbox{simplcolorbox}{
  colback=findingbg, %
  colframe=findingbg, %
  left=0mm, %
  right=0mm, %
  top=0mm, %
  bottom=0mm, %
  boxsep=1mm, %
  arc=0mm, %
  outer arc=0mm, %
  breakable=true,
  before upper={
  \stepcounter{findingcounter}
  \textbf{Finding \thefindingcounter: }}
}

\usepackage{framed}
\newcounter{rqs}[section]
\newenvironment{rqs}{\refstepcounter{rqs}
    \vspace{-1mm}
    \framed
    \noindent \textbf{}
}
{
    \endframed
    \vspace{-1mm}
}

\begin{document}

\title{Embedding Software Intent: Lightweight Java Module Recovery}

\author{Yirui He, Yuqi Huai, Xingyu Chen, Joshua Garcia\\
Department of Informatics, University of California, Irvine\\
Irvine, CA, USA\\
\{yiruih,yhuai,xingyc20,joshug4\}@uci.edu
}

\maketitle

\begin{abstract}
As an increasing number of software systems reach unprecedented scale, relying solely on code-level abstractions is becoming impractical. While architectural abstractions offer a means to manage these systems, maintaining their consistency with the actual code has been problematic.
The Java Platform Module System (JPMS), introduced in Java 9, addresses this limitation by enabling explicit module specification at the language level. JPMS enhances architectural implementation through improved encapsulation and direct specification of ground-truth architectures within Java projects.
Although many projects are written in Java, modularizing existing monolithic projects to JPMS modules is an open challenge due to ineffective module recovery by existing architecture recovery techniques. 
To address this challenge, this paper presents \llm (\llmfull), a novel, lightweight, and efficient approach that recovers \jms from monolithic Java systems using fully-qualified class names. \llm leverages language models to extract semantic information from package and class names, capturing both structural and functional intent.
In evaluations across 20 popular Java projects, \llm outperformed all state-of-the-art techniques in architectural-level similarity metrics while achieving execution times that were 3.99 to 10.50 times faster.
\end{abstract}


\vspace{-.5ex}
\section{Introduction}
\vspace{-.5ex}
Managing large-scale and complex systems through traditional code-level abstractions (e.g., functions, classes, packages) becomes infeasible as codebases reach hundreds of millions of lines. Software architecture provides higher-level abstractions (e.g., components, connectors, interfaces, configurations) to address this complexity~\cite{taylor2008,shaw1996,perry1992foundations}. Unfortunately, ensuring the consistency between architecture-level abstractions and code-level abstractions has been a longstanding problem at the core of software architectural drift and erosion~\cite{taylor2009,perry1992}, which we collectively refer to as \textit{architectural decay}. 

To address architectural decay, \textit{reverse engineering} offers a promising direction by identifying system components and their interrelationships, creating representations at higher levels of abstraction~\cite{chikofsky2002reverse}.
Within the software architecture field, reverse engineering from implementation to high-level abstractions (i.e., software architecture recovery) stands as a critical topic.
Over the past decades, researchers have proposed numerous architecture recovery techniques that extract system architecture from code implementations and reduce the manual effort required to maintain architectural abstractions~\cite{garcia2013comparative,zhang2023software}.
While architecture recovery provides an effective approach for transforming monolithic systems into modular abstractions, two fundamental challenges remain.
(1) Producing highly accurate architecture-recovery techniques faces the difficulty of obtaining accurate ground-truth architectures of projects, resulting in a relatively small number of such architectures being available \cite{garcia2013obtaining,lutellier2017measuring,lutellier2015comparing,garcia2021constructing}. 
(2) Existing recovery techniques have leveraged semantic~\cite{garcia2011enhancing} and structural~\cite{tzerpos2000accd} information, or combinations thereof~\cite{zhang2023software}. However, achieving scalable recovery remains an open challenge, whether measured against ground-truth architectures or encapsulation metrics~\cite{lutellier2015comparing,lutellier2017measuring,garcia2013comparative}.

The introduction of the Java Platform Module System (JPMS) provides a unique opportunity to address the first challenge and, in turn, enables a systematic analysis of the second. Because JPMS encodes developer-created module boundaries directly in source code, it offers high-quality ground truth for large Java projects. By extracting these developer-defined modules, we can construct reliable benchmark architectures at scale. Building on this insight, we introduce \llm (\llmfull), the first technique that extracts Java modules from monolithic systems using only fully-qualified class names, making it lightweight.

Our \textit{\textbf{key insight}} of \llm is that by leveraging LM-based topic models with undersized module repair, we can design a technique to extract Java modules that substantially outperforms existing recovery techniques on a variety of accuracy and encapsulation measures while only needing the fully-qualified name of classes as input. 
The validity of relying solely on class names is supported by our examination of the modularization process of large-scale projects such as JUnit5 and OpenJDK JDK, where over 90\% of source files in modularized versions preserve exact class-name matches with their non-modularized counterparts. This stability is consistent with prior empirical findings that major structural modifications are rare in real-world software evolution~\cite{kim2014empirical,kim2012field}.

Our evaluations show that \llm outperforms existing state-of-the-art architecture recovery techniques by producing \rems that not only more closely match \dms under the measure of a2a score, by 11.46 to 20.63 percentage points (percentage point, pp, is the unit for the arithmetic difference between two percentages), but also exhibit good encapsulation property on Modularization Quality (MQ). 
\llm performs these recoveries 3.99 to 10.50 times faster than existing techniques.

To summarize, this paper makes the following contributions:
\begin{itemize}[leftmargin=*,nosep]
    \item We built a dataset comprising 20 popular open-source GitHub~\cite{githubGitHubLets} projects of varying scales. We extracted the ground-truth architecture for these 20 projects. This dataset can serve as a resource for future research and development.
    
    \item 
    We introduce \llm, the first technique for extracting well-encapsulated \jms that strongly resemble developer-created modules. \llm operates using LMs and only the fully-qualified Java class names of a project as input, making \llm fast and lightweight. 

    \item 
    We evaluated the effectiveness of \llm at extracting \jms that resemble \dms with 20 popular open-source GitHub projects, by comparing it against 4 state-of-the-art architecture recovery techniques. The results demonstrate that \llm outperforms all techniques on \textit{a2a} while achieving a faster speed. 

\end{itemize}

The remainder of the paper is organized as follows: 
\autoref{sec:background} explains the background.
\autoref{sec:research-questions} introduces the research questions.
\autoref{sec:methodology} illustrates the main process of our study and describes \llm in more detail; 
\autoref{sec:result} 
details our analysis and discusses the obtained results;
\autoref{sec:thread-to-validity} describes threats to validity;
\autoref{sec:related-work} outlines related work of the study and \autoref{sec:conclusion} concludes the paper. 

\vspace{-.5ex}
\section{Background} \label{sec:background}
\vspace{-.5ex}
\subsection{Software Architectural Decay}
Software systems have grown increasingly large and more complex, resulting in many legacy software systems, which tend to remain in deployment and use for years or even decades. 
To maintain this growing complexity and scale of legacy software, program analysis of both the static and dynamic variety has been deployed to obtain high-level architectural abstractions from code-level artifacts \cite{koschke2009archrecon, ducasse2009software, garcia2013comparative, lutellier2017measuring, lutellier2015comparing, yan2004discotect, schmerl2005discotect, feiler2012model}. 
Given the massive size of software, the length of time they live nowadays before they are decommissioned, and the continued need to build and maintain such large systems, it is imperative that we consider software constructs and concepts beyond typical code-level abstractions. A major means of handling such large and complex software systems has traditionally been to utilize constructs from software architecture (e.g., components, connectors, interfaces, and configurations)~\cite{taylor2008,shaw1996,perry1992foundations}. 

Unfortunately, ensuring that architecture-level abstractions and code-level abstractions are consistent has been a longstanding problem at the core of \textit{software architectural drift and erosion}~\cite{taylor2009,perry1992}, which we collectively refer to as \textit{architectural decay}. 
Instances of architectural decay manifest themselves in a variety of forms---including architectural tactical vulnerabilities~\cite{da2021understanding}, architectural technical debt \cite{kruchten2012technical}, architectural smells~\cite{fontana2017arcan,smells_csmr,smells_qosa}, architectural hotspot patterns~\cite{xiao2016identifying,cai2017detecting}, and architectural mismatch~\cite{garlan1995icse_mismatch, garlan1995ieee_software_mismatch, garlan2009mismatch}. 

In a study of 1,821 software professionals~\cite{ernst2015ttechdebt}---primarily software engineers and architects---architectural design decisions were the key determinant of technical debt. The study further found that monitoring and tracking architectural decay
was vital and architectural issues were difficult to track since they were often caused years in the past. Other work has shown technical debt, on average, accounts for 36\%~\cite{beskerPriceyBillofTechDebt2017} of wasted developer time, most of which stems from architectural problems.
However, previous work shows that only 38\% of the time developers are discussing the impact of their changes on the architectural structure~\cite{paixao2017developers,paixao2019impact}. Software architectural decay has attracted significant attention and has been subject to extensive study due to the potential for severe consequences it can impose on various industries.
The ramifications of architectural decay can be profound, with some instances even resulting in financial losses for organizations.

\subsection{JDK's Architectural Decay and JPMS}
A major example of recent and major architectural decay that would benefit from architecture recovery is in the Java ecosystem, especially with respect to the Java Development Kit (JDK), which has been described as a ball-of-mud architecture~\cite{openjdkDraftIntegrity, infoqProjectJigsaw}.
More specifically, as the JDK has reached over hundreds of packages, projects built using the JDK have abused access to powerful, unsupported, and dangerous mechanisms in it (e.g., \texttt{sun.misc.Unsafe}~\cite{unsafeAtAnySpeed, mastrangelo2015use}), which gives programs the ability to perform memory allocation and deallocation not subject to garbage collection, resulting in a rigid JDK architecture~\cite{openjdkDraftIntegrity}. 
This ball-of-mud, monolithic architecture and the resulting rigidity have forced JDK engineers to spend so much time maintaining unsupported, even dangerous APIs, that new features and other important forms of maintenance have been unable to be conducted~\cite{openjdk260Encapsulate, openjdk451Prepare, openjdk200Modular, openjdk403Strongly, openjdk396Strongly, openjdk261Module}.

To strongly encapsulate the JDK and other projects and prevent the aforementioned abusive usage, JDK engineers developed the \jpms, which includes \textit{architectural modules} in the Java language. \jms are at a higher level of abstraction and are containers for packages. 
\jms can expose or require their internals using five different types of directives~\cite{ProjectJigsaw, UnderstandingJava9Modules, Java9forProgrammers}---which serve as different forms of provided and required interfaces typically modeled in forward architecture design~\cite{medvidovic2000classification}. 
With the concrete language-level construct, developers can control the visibility and accessibility of code in the module. 
Moreover, \jpms can also provide an architecture (i.e., the content of the modules and the type of inter-communication) for developers to use. Thus, \jpms is getting more and more popular as there are over seven thousand unique Java Modules~\cite{sormurasJPMS}. 

\subsection{Java Module Structure}
\begin{figure}[htb]
\vspace{-2ex}
 \centerline{\includegraphics[width=0.35\textwidth]{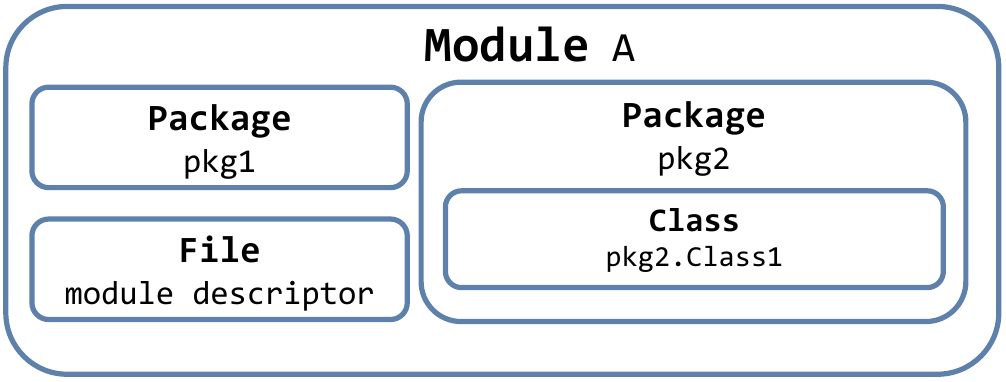}}
        \caption{ \jpms Structure}
        \label{fig:jpms-structure}
\end{figure}

Java modules represent a structural layer above packages, functioning as uniquely identified, reusable units that encapsulate related packages and resources. Each module is defined by a descriptor specifying its name, dependencies, exported packages, service interactions, and reflection permissions. This architecture enhances code organization and inter-component relationship management.
The module system implementation relies on \textit{module directives}, which developers declare in \textit{module declarations} (i.e., \textit{module-info.java}). These declarations compile into \textit{module descriptors} (i.e., \textit{module-info.class}) stored at the module's root \cite{oracleJavaxAELanguage}.

\autoref{fig:jpms-structure} illustrates the structure of module \texttt{A}. The module contains a \textit{module descriptor} that specifies information~\cite{UnderstandingJava9Modules} such as the module name, dependencies on other modules, and packages explicitly exported to external modules. Classes and packages not intended for public use are omitted from the module descriptor. The module follows a hierarchical organization, containing multiple packages (i.e., \texttt{pkg1}, \texttt{pkg2}) that encapsulate fine-grained types. E.g., package \texttt{pkg2} contains class \texttt{Class1} with fully qualified name \texttt{pkg2.Class1}.

\vspace{-.5ex}
\section{Research Questions} \label{sec:research-questions}
\vspace{-.5ex}
In our study, we seek to answer research questions that assess \llm, the first technique designed to recover Java module. 
To that end, we evaluate its effectiveness and efficiency with 5 research questions. 

\vspace{-0.5ex}
\begin{rqs}
\textbf{RQ1}: To what extent do architecture recovery techniques produce architectures that resemble \das? 
\end{rqs}
\vspace{-0.5ex}

In prior work, a high-quality software architecture that includes a mapping of an architectural module to implementation-level entities (e.g., classes, files, or packages) often requires confirmation by a project's developers or architects~\cite{garcia_obtaining_2013,lutellier2015comparing,lutellier2017measuring}. With the increasing adoption of Java modules in popular Java projects, we are able to use such projects and their \dms as ground truth. 
More specifically, Java modules recovered using architecture recovery techniques should resemble \dms. 
As a result, for RQ1, we measure the extent to which \rems resemble the \dms at both the system and module levels.

\vspace{-0.5ex}
\begin{rqs}
\textbf{RQ2}: How well-encapsulated are modules produced by recovery techniques? 
\end{rqs}
\vspace{-0.5ex}

\jpms provides mechanisms to strongly encapsulate Java modules and make them reusable \cite{UnderstandingJava9Modules}. 
Architectural encapsulation is often measured in terms of a software entity's coupling and cohesion \cite{mancoridis1999bunch,garcia2013comparative,lutellier2017measuring,lutellier2015comparing,tzerpos2000accd, bouwers2014quantifying}. 
As a result, for RQ2, we assess the extent to which \rems exhibit minimum coupling and maximum cohesion using architectural encapsulation metrics. 
Studying RQ2 can shed some insight into the quality of \rems.

\vspace{-0.5ex}
\begin{rqs}
\textbf{RQ3}: What is the runtime efficiency of the state-of-the-art architecture recovery techniques under study compared to \llm?
\end{rqs}
\vspace{-0.5ex}

Ideally, a recovery technique should be fast when constructing a high-quality architecture. 
Thus, we measure and compare the time each recovery technique takes to recover the architectures of the projects we study.

\vspace{-0.5ex}
\begin{rqs}
\textbf{RQ4}: To what extent does each component in \llm contribute to its effectiveness?
\end{rqs}
\vspace{-0.5ex}

To study how each component of \llm contributes to the overall Java module recovery process, we conduct an ablation study based on the key components of \llm.

\vspace{-0.5ex}
\begin{rqs}
\textbf{RQ5}:  To what extent does \llm's effectiveness vary when using different levels of input granularity?
\end{rqs}
\vspace{-0.5ex}

To assess input sensitivity, we evaluate \llm's performance across three distinct input configurations: (1) expanded input comprising complete class implementations, (2) baseline input using fully-qualified class names, and (3) reduced input containing only package names. This systematic variation allows us to quantify the impact of input granularity on recovery effectiveness.

\vspace{-.5ex}
\section{Study Methodology and \textit{C\MakeLowercase{lass}LAR}}\label{sec:methodology}
\vspace{-.5ex}

In this section, we discuss the research methodology we follow to answer the research questions described in the previous section. 
First, we cover our selection criteria for the projects we study and the compilation process we utilize for the projects. 
Second, we describe how we identify the ground-truth architectures for those projects, i.e., \das.
Lastly, we (1) introduce the selection criteria for the state-of-the-art architecture recovery techniques we utilize as baselines for comparison with \llm and (2) describe \llm in more detail.

\subsection{Selection of Java Projects}

We collected JPMS projects from GitHub~\cite{githubGitHubLets}, a large and popular platform for open-source software.
Our criteria for selecting projects from GitHub are as follows: The project must (1) have a sufficient number of modules (5 or more) to ensure recovering the architecture is not trivial, (2) use Java as the primary programming language, (3) be sufficiently popular, i.e., starred by 200 or more users, 
and (4) be still actively maintained. 
These selection criteria ensure that our selected projects are widely used and that our study results are generalizable.
\autoref{tab:subjects_overview} shows an overview of all 20 projects we collected that satisfy the aforementioned criteria.

During the compilation process of a project, we adhere to three rules to ensure our data is most up-to-date and reproducible: (1) We select the latest release version of the project that can be compiled; 
(2) we use the specified JDK version for compilation; and 
(3) in cases where the version of the JDK is not specified, we use JDK 11, which is the oldest stable version that succeeds Java 9. Moreover, if the specified JDK is older than Java 9, we also adopt JDK 11. 

\subsection{Determining Developer-Created Modules} \label{sec:determing-dcm}

Developer-created modules often serve as ground truth for determining which code-level entities (i.e., packages and classes) should constitute each module~\cite{lutellier2017measuring,garcia2013comparative}.
However, prior to Java including architectural modules in its language, finding or constructing such ground-truth architectures was challenging, resulting in few such architectures existing \cite{garcia2013obtaining,garcia2021constructing}.

To determine the code-level entities within JPMS modules, we rely on the fact that module declarations are specified in a \textit{module description file} (i.e., \textit{module-info.java}) and are compiled into a \textit{module descriptor} (i.e., \textit{module-info.class}). However, module descriptors primarily define module dependencies and exported packages rather than explicitly listing constituent classes.
We observed two types of structures that store module descriptors from our selected projects: 
The first type of structure is one in which the root directory contains a \textit{module descriptor}; 
the second type is a JAR (i.e., Java ARchive) that contains a \textit{module descriptor}. 
For the first type of structure, all the code-level entities located within the module root directory are entities contained within a \dm defined by the module descriptor in the root directory. 
For the second type of structure, all entities located within the JAR file are contained within the module described by the JAR's module descriptor.

\subsection{Recovering Java Modules}
\label{sec:methodology-module-structure}

\subsubsection{Baseline Recovery Techniques} 
\label{sec:methodology:recovering_modules:baselines}

We selected four baseline techniques with historically good performance across multiple studies~\cite{garcia2021forecasting,le2015empirical,lutellier2015comparing,lutellier2017measuring,zhang2023software,teymourian2020fast}, i.e., ACDC \cite{tzerpos2000accd} and ARC~\cite{garcia2013comparative, garcia2011enhancing}, SARIF~\cite{zhang2023software} and FCA~\cite{teymourian2020fast}.

These techniques have been shown to have a high accuracy when compared to manually recovered architectures obtained with the actual architects of widely-used non-JPMS software systems (e.g., Hadoop) \cite{garcia2013comparative,zhang2023software}. 
Besides exhibiting high accuracy on non-JPMS projects, they utilize different kinds of information to recover architectures:
\begin{itemize}[leftmargin=*]
    \item ACDC~\cite{tzerpos2000accd} recovers software architectures based on a variety of patterns mainly related to the static control- or data-flow dependencies of code-level entities.
    \item ARC leverages a topic model, i.e., Latent Dirichlet Allocation (LDA)~\cite{blei2003latent}, to recover architectures. A topic model consists of \textit{topics}, i.e., probability distributions of words (e.g., identifiers and comments in code), and treats code-level entities (e.g., classes) as \textit{documents}, i.e., probability distributions over topics, to recover software architectures.
    \item SARIF~\cite{zhang2023software} is the most recent technique that recovers architecture based on dependencies, code text, and folder structure. 
    However, one of its inputs—absolute file paths—implicitly encodes ground-truth Java module information. As described in \autoref{sec:determing-dcm}, packages belonging to the same module are typically colocated in a folder with the module descriptor at the root, so developers often organize the source code in the same way. This invalidating SARIF for Java module recovery. 
    To compare against the tool, we modified the technique to avoid using the file path information. Thus, the SARIF referred to in the evaluation section is the modified and valid version.
    \item FCA~\cite{teymourian2020fast} aimed to maximize the cohesion within and minimize the coupling between clusters. 
    Although FCA uses similar input types as ACDC, it achieves lower performance in terms of ground-truth architecture resemblance~\cite{teymourian2020fast} and was not originally designed for Java projects. However, we include FCA due to its demonstrated effectiveness on a Java-based system (i.e., OODT)~\cite{zhang2023software}.
\end{itemize}

Unlike those techniques, \llm only needs the fully-qualified class name as input and utilizes an LM to build the topic model and recover the architecture.

\par
\begin{table}
    \caption{Subject Project Overview }
    \centering
    \scriptsize
    \begin{tabular}{p{.17\textwidth}|r|r|r|r}
    \toprule
    \hline
        \rowcolor[RGB]{198,208,230}  \textbf{Project Name} & \textbf{\# Stars} & \textbf{\# Modules} & \textbf{\# File} & \textbf{\# LOC} \\ \hline
         \textbf{OpenJDK JDK} & 14,713 & 69 & 50,860 & 6,334,312 \\ 
        \rowcolor[RGB]{224,230,242} \textbf{DL4J} & 12,715 & 42 & 5,083 & 1,038,356 \\ 
        \textbf{JUnit5} & 5,529 & 9 & 1,377 & 95,938 \\ 
        \rowcolor[RGB]{224,230,242}  \textbf{LWJGL3} & 4,037 & 46 & 5,100 & 679,869 \\ 
         \textbf{SLF4J} & 2,055 & 7 & 258 & 13,682 \\ 
        \rowcolor[RGB]{224,230,242} \textbf{Speedment} & 2,039 & 40 & 3,045 & 177,327 \\ 
         \textbf{FlatLaf} & 2,027 & 9 & 433 & 69,573\\ 
        \rowcolor[RGB]{224,230,242}  \textbf{BC-Java} & 1,827 & 6 & 6,085 & 874,638 \\ 
         \textbf{OpenJDK Loom} & 1,592 & 68 & 50,764 & 6,333,448 \\ 
        \rowcolor[RGB]{224,230,242}  \textbf{Apache Lucene} & 1,378 & 36 & 5,301 & 841,199 \\ 
         \textbf{Ebean ORM} & 1,343 & 44 & 3,705 & 218,197 \\ 
        \rowcolor[RGB]{224,230,242} \textbf{Microsoft GCToolKit} & 1,158 & 5 & 274 & 20,489\\ 
         \textbf{JetBrains Runtime} & 745 & 69 & 51,049 & 6,371,075\\ 
        \rowcolor[RGB]{224,230,242}  \textbf{Cache2k} & 672 & 5 & 546 & 51,608 \\ 
         \textbf{CalendarFX} & 656 & 8 & 293 & 36,402 \\ 
        \rowcolor[RGB]{224,230,242} \textbf{JFXtras} & 588 & 11 & 615 & 57,969 \\ 
         \textbf{Tinylog} & 545 & 13 & 469 & 47,122 \\ 
        \rowcolor[RGB]{224,230,242}  \textbf{Sejda SDK} & 422 & 7 & 586 & 29,925 \\ 
         \textbf{Ikonli} & 414 & 62 & 415 & 61,213\\ 
       \rowcolor[RGB]{224,230,242}  \textbf{Apache Derby} & 290 & 23 & 2,910 & 732,208 \\ 
       \hline
    \bottomrule
    \end{tabular}
    \label{tab:subjects_overview}
    \vspace{-4ex}
\end{table}

\subsubsection{\llm} \label{sec:llm-recovery-technique} 
\begin{figure}[htb]
 \centerline{\includegraphics[width=0.5\textwidth]{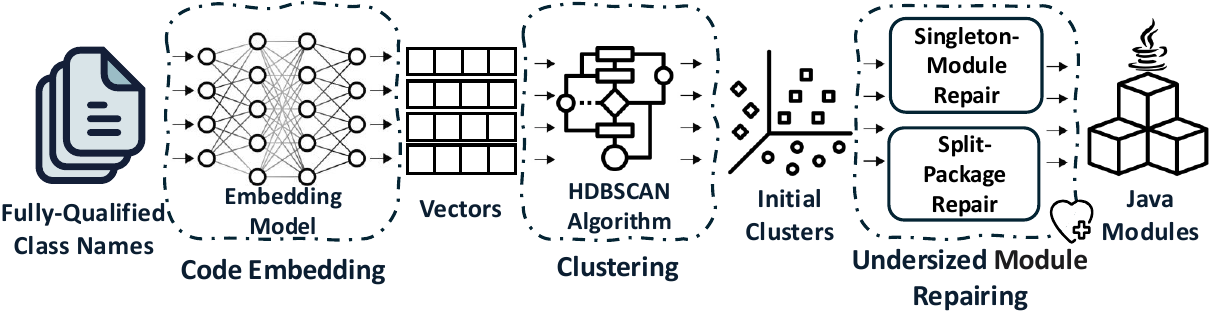}}
        \caption{\llm Recovery Process  
        }
        \label{fig:llm-technique-process}
\end{figure}

The high-level recovery process of \llm is shown in \autoref{fig:llm-technique-process}.
\llm begins by extracting fully-qualified class names from software systems, utilizing these names as input for the embedding model (i.e., LM). 
The embedding model is an algorithm trained to encapsulate information into dense representations in a high-dimensional space, aiming to transform the textual inputs into numerical vectors. 
After the creation of these vectors, \llm employs an unsupervised clustering algorithm to organize the vectors into distinct clusters, where each vector represents a class and each cluster represents a Java module. 
Unfortunately, clustering algorithms can not prevent the formation of singleton clusters (i.e., Java modules containing only one class) or the split-package issue~\cite{kothagal2017modular} (i.e., a named Java package distributed across different modules).
To address these limitations, \llm includes an undersized module repair step, which involves merging singleton clusters (i.e., clusters containing only one Java class) with larger clusters, and grouping classes from the same package into the same Java modules. 
This section further details the design and implementation specifics of our approach. We employ default hyper-parameter configurations throughout our implementation unless explicitly stated otherwise.

\vspace{-2ex}

\llm uses the fully-qualified class name for each class (e.g., \texttt{pkg2.Class1} in \autoref{fig:jpms-structure}), as opposed to static dependencies used in prior work (e.g., ACDC) or the code of source-level entities (e.g., identifiers and comments of classes) used by other NLP-based architecture recovery techniques (e.g., ARC). 
As our evaluation will show, inputting the entire code of a class actually worsens the accuracy and encapsulation of \llm's recovered architectures, suggesting a fully-qualified class name is all the input \llm needs.

\noindent 
\textbf{\textit{$\bullet$ Code Embedding.}} 
Our approach represents each fully-qualified class name as a dense semantic vector using a pretrained code embedding model. The model encodes lexical and structural information from class identifiers and package hierarchies, positioning semantically related classes proximally in the embedding space. These vectors serve as inputs for unsupervised clustering to recover module boundaries in a data-driven manner.

\noindent 
\textbf{\textit{$\bullet$ Clustering}.} \llm employs HDBSCAN~\cite{mcinnes2017hdbscan}, a hierarchical density-based clustering algorithm.
To address the well-known challenges of high-dimensional data clustering~\cite{assent2012clustering}, 
\llm pre-process the vectors by performing dimensionality reduction using UMAP (Uniform Manifold Approximation and Projection)~\cite{mcinnes2018umap},
which preserves both local and global data structures~\cite{grootendorst2022bertopic, guo2018adult}. Following established best practices~\cite{bertopicumap}, we utilize cosine similarities rather than Euclidean distance metrics for the dimensional reduction process, as the latter can be problematic for high-dimensional spaces~\cite{domingos2012few, beyer1999nearest}.

\begin{figure}[htb]
\vspace{-1ex}
 \centerline{\includegraphics[width=0.48\textwidth]{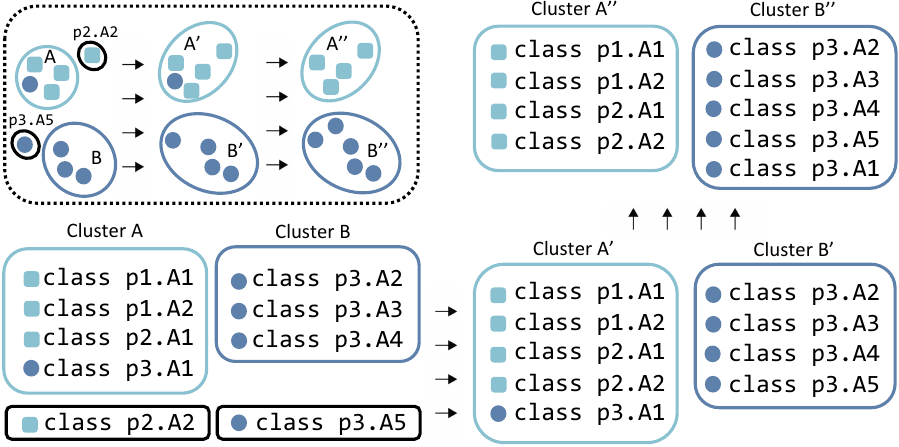}}
        \caption{Undersized Modules: A clustering with singleton modules (p2.A2 and p3.A5) and split packages p2 and p3.}
        \label{fig:undersize-module-repair-example}
\end{figure}

\noindent 
\textbf{\textit{$\bullet$ Undersized Module Repairing.}}
To repair undersized modules, \llm eliminates (1) \textit{singleton modules} (i.e., modules that only have 1 Java class) and (2) any \textit{split package}~\cite{kothagal2017modular} (i.e., a named Java package distributed across different modules). 
Singleton modules resulting from software architecture recovery have been known to be problematic~\cite{shtern_tzerpos_clustering_methods_for_se,antequil+lethbride_exp_clustering_for_remodularization} and do not resemble ground-truth components or modules~\cite{garcia2013obtaining}.
\autoref{fig:undersize-module-repair-example} shows example singleton modules as represented by a cluster with only class $p2.A2$ and another cluster with only class $p3.A5$.
Prior to Java 9 and JPMS, split packages can lead to unpredictable or erroneous behavior due to multiple classes or packages with the same name being loaded from the Java classpath \cite{parlog2019java}. 
Starting with Java 9, split packages are detected by the Java compiler and result in a compile-time error \cite{parlog2019java}. 
In \autoref{fig:undersize-module-repair-example}, package $p2$ is split across cluster A and a singleton module $p2.A2$, and package $p3$ is split across cluster A, B, and a singleton module.

To eliminate singleton modules, \llm employs c-TF-IDF representations~\cite{grootendorst2022bertopic,ctfidf}
to assign each singleton module to its best-matching non-singleton module. c-TF-IDF extends the traditional TF-IDF metric~\cite{aizawa2003information} used in topic modeling~\cite{grootendorst2022bertopic} by operating at the cluster (i.e., Java module) level rather than the document (i.e., Java class) level. The process first concatenates all Java classes within a cluster and computes their adjusted TF-IDF values. As shown in \autoref{fig:undersize-module-repair-example}, this singleton-module repair reassigns $p2.A2$ to cluster A (forming $A'$) and $p3.A5$ to cluster B (forming $B'$).
Subsequently, \llm resolves split packages by relocating all Java classes from the same package to the cluster containing the majority of that package's classes. In \autoref{fig:undersize-module-repair-example}, the repair step moves $p3.A1$ to cluster $B'$, yielding the final clusters $A''$ and $B''$.

The recovered modules are finalized after the undersized modules have been repaired.

\newcounter{rq_counter}[section]
\section{Evaluation Setup} \label{sec:result}

To answer five research questions we proposed in \autoref{sec:research-questions}, we ran all our experiments on a workstation with 2 AMD EPYC 7551 32-core processors with 512 GB of RAM. 
We allotted a 12-hour period for each recovery technique to reconstruct the architecture of each given project, which is about 6 times longer than the time allotted for recovery techniques in previous work~\cite{lutellier2017measuring}. 
Due to the non-determinism of \llm, we ran each experiment 3 times and took the average score as our final score.

To prevent data leakage, we selected an open-source embedding model trained exclusively on publicly available datasets.
We implemented a systematic model selection process leveraging the Hugging Face platform~\cite{neulabcodebertjava}, which hosts a comprehensive repository of state-of-the-art language models. Our selection criteria prioritized Java-specific models, given our study's focus on Java systems, and ranked candidates based on their community adoption rates as measured by monthly download metrics.
Based on this selection process, we obtained CodeBERT-java~\cite{neulabcodebertjava}. CodeBERT-java extends the base CodeBERT~\cite{feng2020codebert} architecture with additional training on Java-specific corpora, enabling more precise representations for Java-related tasks such as documentation generation.
While we focus on CodeBERT-Java to ensure reproducibility and avoid data leakage concerns, our approach is generalizable to other embedding models, including more advanced proprietary models such as OpenAI embeddings~\cite{openaiOpenAIPlatform}.

\section{Result and Analysis}
\refstepcounter{rq_counter}
\subsection{ RQ\arabic{rq_counter}: Recovered Modules Resemblance } \label{sec:rq2}

\subsubsection{Evaluation Metrics}

To determine whether the selected architecture recovery techniques can produce architectures similar to \das, we used the \textit{architecture-to-architecture (a2a)} score, which has been used by previous work on architecture recovery~\cite{le2015empirical, garcia2013comparative, lutellier2015comparing, zhang2023software, schmitt2020arcade} to study architecture similarity at the system level.
\textit{a2a} achieves architecture-level comparison by measuring the smallest edit distance from one architecture to another. The \textit{a2a} score is calculated as follows:
\vspace{-1ex}
\begin{equation}
\begin{aligned}
&a2a(A_{i}  ,  A_{j})=(1- \frac {mto(A_{i},A_{j})}{aco(A_{i})+aco(A_{j})}  )  \times  100\%   \nonumber
\end{aligned}
\end{equation}
\noindent where $A_i$ and $A_j$ are two different software architectures, $mto(A_i, A_j)$ is the minimum number of operations needed to transform architecture $A_i$ into $A_j$, $aco(A)$ is the number of operations needed to construct architecture $A$ from a ``null'' architecture. Specifically:
\vspace{-1.5ex}
\begin{equation}
\begin{aligned}
mto(  A_{i}  &,  A_{j}  )=remC(  A_{i},  A_{j}  )+addC(  A_{i}  ,  A_{j}  )\\ &+remE(  A_{i}  ,  A_{j}  )
+addE(  A_{i}  ,  A_{j}  )+movE(  A_{i}  ,  A_ {j} ) \nonumber
\end{aligned}
\end{equation}
\vspace{-2ex}
\begin{equation}
\begin{aligned}
aco(  A_{i}  )=addC(  A_{\phi}  ,  A_{i}  )+addE(  A_ {\phi}  ,  A_{i}  )+movE(A_{\phi}  ,  A_{i}) \nonumber
\end{aligned}
\end{equation}

\noindent where $A_\phi$ is a null architecture and the operations comprise entity additions ($addE$), removals ($remE$), and moves ($movE$) between modules, along with module additions ($addC$) and removals ($remC$).

To study the similarity of recovered and developer-created architectures at a finer-grained level, i.e., the module/cluster level, we selected two well-established metrics: homogeneity score (\hscore) and completeness score (\cscore)~\cite{rosenberg2007v}. 
An alternative metric for module-level similarity is the \textit{cluster-to-cluster (c2c)} score~\cite{le2015empirical, garcia2013comparative, lutellier2015comparing}. However, as shown in previous work \cite{zhang2023software}, \ctoc requires selecting one or more thresholds that are difficult to determine and, even when selected, often give unstable and unreasonable results, i.e., fluctuates excessively and in a non-intuitive manner.
As a result, we preclude using \ctoc and use the \hscore and \cscore instead.

A recovery technique achieves a higher \hscore for a specific \rem when it exclusively assigns classes from a single \dm to it. Conversely, a recovery technique achieves a higher \cscore for a specific \rem when all its constituent classes belong to the same \dm.

Assume we have a set of 
\dms $C = \{c_i \mid i = 1,...,n\} $ and a set of \rems $K = \{k_i \mid i = 1,...,m\} $. 
Also, $a_{ij}$ represents the number of classes that are members of \dms $c_i$ and elements of \rems $k_j$. $N$ is the total number of Java classes. 
The scores are calculated using the following methods:
\vspace{-1ex}
\begin{equation}
\begin{aligned}
\notag
    \textrm{\hscore}=1-\frac{H(C|K)}{H(C)}  &\quad 
    \textrm{\cscore}=1-\frac{H(K|C)}{H(K)}
\end{aligned}
\end{equation}
\vspace{-1ex}
\noindent where 
\vspace{-3ex}
\begin{equation}
\begin{aligned}
\notag
    H(C|K)=-\sum_{k=1}^{|K|}{\sum_{c=1}^{|C|}\frac{a_{ck}}{N}log{\frac{a_{ck}}{\sum{_{c=1}^{|C|}a_{ck}}}}} \\
    H(K|C)=-\sum_{c=1}^{|C|}{\sum_{k=1}^{|K|}\frac{a_{ck}}{N}log{\frac{a_{ck}}{\sum{_{k=1}^{|K|}a_{ck}}}}}\\
    H(C) = -\sum_{c=1}^{|C|}\sum_{k=1}^{|K|}{\frac{a_{ck}}{n}log\frac{\sum_{k=1}^{|K|}a_{ck}}{n}} \\
    H(K) = -\sum_{k=1}^{|K|}\sum_{c=1}^{|C|}{\frac{a_{ck}}{n}log\frac{\sum_{c=1}^{|C|}a_{ck}}{n}}
\end{aligned}
\end{equation}
\vspace{-1ex}

\noindent When there is only one \dm (i.e., $H(C) = 0$), we define \hscore to be 1. Similarly, if any \rea only contains one \rem (i.e., $H(K)= 0$), we define \cscore to be 1. 

\subsubsection{Results}
The results for each recovered architecture are presented in \autoref{table:merged-metrics}.
Both ACDC and ARC failed to recover architectures for 3 subjects 
as ACDC timed out and ARC faced memory errors during topic model generation. 
Overall, \llm performs the best in successfully recovering all architectures without any failure. It also achieves the highest individual and average a2a scores, leading to 15 out of 20 \reas and 11.46 pp higher than the 2nd highest-ranked recovery technique.

\vspace{-1ex}
\begin{simplcolorbox}
For a2a's more coarse-grained analysis, \autoref{table:merged-metrics} indicates that \llm can generate architectures resembling the \da with an average similarity higher than 78\% and achieves the highest average a2a score of 85.78\%, which is 11.46 pp greater than the next-best technique, ACDC. 
\llm also successfully recovered 15 out of 20 architectures most similar to the \da. 
\end{simplcolorbox}

\begin{table}[]
\caption{Average Similarity and Encapsulation Metrics Across All Techniques}
\label{table:merged-metrics}
\centering
\scriptsize
\begin{tabular}{l|c|c|c|c}
\toprule\hline
\rowcolor[RGB]{198,208,230}
\textbf{Technique} & \textbf{a2a} & \textbf{\cscore} & \textbf{\hscore} & \textbf{MQ} \\ \hline
ACDC  & 70.73\% & 35.60\% & 59.07\% & 8.11\% \\
\rowcolor[RGB]{224,230,242} ARC   & 65.15\% & 23.82\% & 46.08\% & 0.90\% \\
SARIF & 74.32\% & 36.88\% & 30.74\% & \textbf{19.28\%} \\
\rowcolor[RGB]{224,230,242} FCA   & 73.65\% & 49.46\% & \textbf{77.55\%} & 3.96\% \\
\llm  & \textbf{85.78\%} & \textbf{56.27\%} & 77.44\% & 15.61\% \\
\hline
\bottomrule
\end{tabular}
\vspace{-3ex}
\end{table}

At the module level, we computed the \hscore and \cscore. 
\autoref{table:merged-metrics} depicts the average \cscore and \hscore results respectively for all five recovery techniques across the subjects under study. Similar to the system-level a2a results, \llm achieves the highest average \cscore and demonstrates comparable \hscore performance to FCA. However, FCA's \hscore may be inflated since it produces an average of 850.5 singleton modules per project, which can misleadingly boost this metric.

\vspace{-1ex}
\begin{simplcolorbox}
\llm achieves a 56.27\% \cscore---which is 6.81 pp greater than the second best recovery technique's 49.46\% \cscore. 
Furthermore, \llm effectively recovered Java modules most similar to the \dms in terms of \cscore in half of the subjects within our dataset. 
This result shows that \llm recovers Java modules that resemble \dms with greater efficacy than existing recovery techniques.
\end{simplcolorbox}
\vspace{-1ex}

ACDC, ARC, FCA, and \llm demonstrate higher average \hscore values compared to their respective \cscore values, indicating these techniques successfully cluster classes from individual \dms into corresponding \rems, but comparatively fall short of achieving complete class coverage within each \rem. This disparity between \hscore and \cscore suggests that while the techniques effectively identify class relationships within modules, they struggle to capture the comprehensive module boundaries.
For example, \llm achieves an average of 77.44\% \hscore, suggesting it can recover an architecture by grouping a majority of related classes into a single \rem in a manner similar to the corresponding \dms. However, \llm's \cscore is lower than its \hscore, implying that, although a majority of classes from the same \rem are from the same \dm, there are instances from one \dm placed into different \rems.

During our manual inspection, we observed that each \rea produces more \rems than \dms on average: ACDC generates 96 \rems, ARC generates 372 \rems, FCA generates 1365 \rems, and \llm generates 80 \rems, all compared to 29 \dms on average.
When this occurs, \dms are more likely scattered across multiple \rems, instead of multiple \dms being merged into one \rem. This is consistent with our finding that \hscore generally exceeds \cscore.

Similarly, for each individual subject, when the number of \rems is smaller than the number of \dms, \rems are likely to have classes from multiple \dms.
For example, Ikonli's (recovered by \llm) \hscore is 33.28\%, as opposed to its \cscore, which is 100\%.
This discrepancy arises from the fact that while Ikonli, when recovered by \llm, contains only 3 \rems, these encapsulate all Java classes located in Ikonli's 7 \dms. 
As no \dm's content is scattered across multiple \rems, the \cscore reached 100\%. 

\vspace{-1ex}
\begin{simplcolorbox}
Modules recovered by most techniques (i.e., ACDC, ARC, FCA, \llm) typically contain classes from a single \dm, and classes from one \dm are often separated into different \rems, resulting in \hscores that are, on average, 20 pp higher than \cscores across all subjects.
\end{simplcolorbox}
\vspace{-1ex}

\vspace{-0.2ex}
On the contrary, SARIF, which has 18 \rems on average, exhibits distinct characteristics compared to the other 4 techniques, primarily in its tendency to group multiple classes from different \dms into the same \rems, especially for large-scale systems. For example, JDK recovered by SARIF only contains 12 \rems while the ground truth has 69 \dms. This clustering behavior is reflected in its metrics, where the average \cscore (36.88\%) exceeds the average \hscore (30.74\%).  Notably, in 18 of 20 \rea instances, its \cscore surpasses the corresponding \hscore, demonstrating this systematic pattern. Despite this consistent behavioral pattern, SARIF's performance remains suboptimal, failing to achieve satisfactory scores in both \hscore and \cscore metrics compared to \llm.

\refstepcounter{rq_counter}
\subsection{RQ\arabic{rq_counter}: Encapsulation Quality of Java Modules }
\label{sec:rq3}

\subsubsection{Evaluation Metrics}
This RQ compares architectures recovered using five architecture recovery techniques, i.e., ACDC, ARC, SARIF, FCA and \llm, with \das, in terms of encapsulation metrics. 
We adopted the following metric which are designed for measuring encapsulation at an architectural level:
\begin{itemize}
    \item \textit{Modularization Quality} (\textbf{MQ})~\cite{ mancoridis1999bunch, schmitt2020arcade} is a measure of a system's coupling and cohesion, and is calculated as $M Q=\sum_{i=1}^k C F_i$. $k$ is a partition's number of modules, $C F_i$ is the \textit{cluster factor} of module $i$, representing its coupling and cohesion, and is defined as: $CF_i= \frac{2 \mu_i}{2 \mu_i+\sum_{\substack{j=1, j \neq i}}^k\left(\varepsilon_{i, j}+\varepsilon_{j, i}\right)}$, where $\mu_i$ is the number of edges within the module, which measures cohesion. $\varepsilon_{i, j}$ is the number of edges from module $i$ to module $j$, and $\varepsilon_{j, i}$ is the number of edges from module $j$ to module $i$. A higher value for MQ denotes enhanced cohesion and reduced coupling, signifying an optimized, well-structured system architecture.
\end{itemize}

\subsubsection{Result}
\autoref{table:merged-metrics} depicts the MQ results for all recovery techniques and \das: 
In the context of MQ architecture recovery, both \llm and SARIF demonstrate superior performance, each excelling in approximately half of the 20 evaluated projects. While \llm shows a marginal decrease (3.67 pp) in average score compared to SARIF, it maintains a substantial lead (7.50 pp) over the remaining techniques.

As noted in RQ1, SARIF tends to produce fewer \rems, especially for large-scale systems like OpenJDK JDK, OpenJDK Loom, JetBrains Runtime, and DL4J. This tendency inflates SARIF's MQ scores by at least 9.67 pp above \llm's in these 4 projects. More specifically, SARIF recovered only 12 \rems for OpenJDK JDK (69 \dms) while \llm recovered 324 \rems. Reducing modules potentially boost MQ scores since they increase cohesion and decrease coupling. Despite this metric bias favoring SARIF, \llm achieves comparable performance with scores only 3.67 pp lower.

\vspace{-1ex}
\begin{simplcolorbox}
\label{finding:mq}
In MQ architecture recovery, \llm demonstrates comparable effectiveness to SARIF while significantly outperforming other automated recovery techniques, even though SARIF inflates MQ scores by creating fewer large modules. Notably, \llm's performance metrics are, on average, at least twice as high as those of other automated recovery techniques.
\end{simplcolorbox}

\begin{simplcolorbox}
\label{finding:mq}
\llm demonstrates superior effectiveness in Java module recovery across both similarity metrics (a2a, \hscore, and \cscore) and encapsulation metrics (MQ) compared to 4 state-of-the-art techniques (i.e., ACDC, ARC, SARIF, FCA). It outperforms the second-best recovery technique by substantial margins in a2a (11.46 pp) and \cscore (6.81 pp).  While the performance differential with FCA and SARIF is marginal in \hscore and MQ, \llm maintains substantial advantages over the remaining techniques, with improvements of 18.37 pp in \hscore and 7.5 pp in MQ.
\end{simplcolorbox}
\vspace{-1ex}

\refstepcounter{rq_counter}
\subsection{RQ\arabic{rq_counter}: Runtime Efficiency of the Recovery Techniques}

In this RQ, we investigate the runtime efficiency of \llm and the other 4 selected architecture recovery techniques by measuring their execution time on the 20 projects we collected from GitHub.

\begin{figure}
   \centerline{\includegraphics[width=0.45\textwidth]{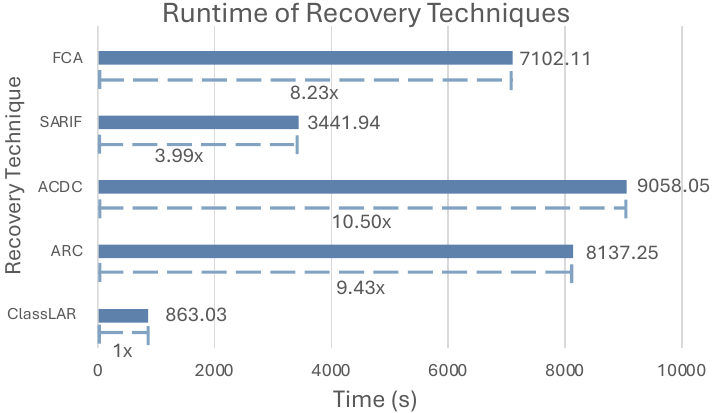}}
        \caption{Runtime Evaluation 
    }
        \label{fig:llm-runtime}
        \vspace{-4ex}
\end{figure}

All 5 techniques successfully recovered architectures for 17 out of 20 projects. 
However, for the remaining 3 projects, i.e., JetBrains Runtime, OpenJDK Loom, and OpenJDK JDK, \llm spent 3,448, 3,605, and 3,628 seconds to recover their architectures, while ACDC failed to complete within the 12-hour time budget and ARC failed due to memory errors. 
\autoref{fig:llm-runtime} displays the average time needed by each of the 5 techniques to recover an architecture across our projects. Since ACDC and ARC failed to recover architectures for 3 of the projects within the time budget, we assume the time needed by those techniques is at least 12 hours and obtain the lower bound of their average runtime.

It is worth emphasizing that \llm's recovery process efficiency benefits from relying solely on class naming conventions from source files. Using Java fully-qualified class names also facilitates effective and generalizable clustering through meaningful semantic patterns.

\vspace{-1ex}
\begin{simplcolorbox}
\llm is capable of recovering architectures for large-scale projects. On average, \llm takes 863.03 seconds to recover the architectures of 20 projects used in this evaluation, which is at least 3.99 times faster when compared against the other 4 techniques.
\end{simplcolorbox}
\vspace{-1ex}

\subsection{RQ4: Ablation Study of \llm}

To further learn each component's effect within \llm and gain greater insight into those key components, we designed an ablation study targeting the major components of \llm, which are its \textbf{code embedding} and \textbf{undersized module repair (UMR)}.

\subsubsection{Experiment Design}

The embedding model is the initial critical component within \llm's process. 
For this study, we selected a deep learning-based natural language processing model, CodeBERT-java~\cite{huggingfaceNeulabcodebertjavaHugging}, which is pre-trained in natural language and programming language, and then fine-tuned on Java programming language datasets.
Despite its typical characterization as a \textit{black box} due to a lack of explainability of such a deep learning model~\cite{zini2022explainability}, the embedding phase is critical as subsequent steps are contingent upon the word embeddings it generates. 
As an alternative to the deep learning model, we employ LDA (recall \autoref{sec:methodology:recovering_modules:baselines}) for producing word embeddings to assess the impact on the overall recovery process. 
We used this alternative because ARC, another natural language-based recovery technique, used LDA to extract a topic model~\cite{garcia2011enhancing}. 
It is worth noting that while obtaining tokens from original input, we split words using camel case like in previous work~\cite{garcia2011enhancing}.
For UMR, we assess the necessity of this phase by removing it for this RQ.

\subsubsection{Result Analysis}
For this RQ, we utilize previously introduced similarity (i.e., a2a, \hscore, and \cscore) and encapsulation (i.e., MQ) metrics. \autoref{table:ablation-study} summarizes the average accuracy changes across 20 subjects when modifications are made to the original \llm setup.

\begin{table}[hbpt]
\centering
\scriptsize
\vspace{-1ex}
\caption{Ablation Study: The best score is \textbf{bolded}. 
- and + indicate value reduction or increase, respectively, for a metric}
\vspace{-1ex}
\begin{tabular}{r|r|r|r}
\toprule
\hline
\rowcolor[RGB]{198,208,230}
\textbf{Metric}   & \textbf{\llm} & \textbf{LDA} & \textbf{UMR} \\
\hline
\textbf{a2a    }         & \textbf{85.78\%}                       & -0.36 pp  & -4.34 pp  \\
\rowcolor[RGB]{224,230,242}
\textbf{\hscore } & 77.44\%                     & -39.86 pp  & \textbf{+0.85 pp}
\\
\textbf{\cscore }    & \textbf{56.27\%}  & -9.24 pp  & -11.49 pp \\
\rowcolor[RGB]{224,230,242}
\textbf{MQ   }          & \textbf{15.61\%}                       & -2.08 pp  & -10.13 pp 
\\ \bottomrule                   
\end{tabular}
\label{table:ablation-study}
\vspace{-2ex}
\end{table}

\vspace{-1ex}
\begin{simplcolorbox}
Removing UMR from \llm worsens the encapsulation of its recovered architectures by 10.13 pp, the system-level architecture resemblance to \das (a2a) by 4.34 pp, and the module-level \cscore by 11.49 pp, while only gaining 0.85 pp in terms of \hscore.
\end{simplcolorbox}
\vspace{-1ex}

This loss in terms of the \hscore likely occurs because UMR primarily helps singleton clusters integrate into appropriate larger clusters. Since each singleton cluster has an \hscore of 1---each cluster containing only one Java class from a single \dm---merging these singletons usually results in a lower average \hscore compared to when they remain isolated. 
However, given that the \hscore and \cscore tend to pull in opposite directions, the very slight loss in \hscore (0.85 pp) is a small price to pay for the significant gain in \cscore (11.49 pp).

\vspace{-1ex}
\begin{simplcolorbox}
Replacing the LM embedding model with a traditional LDA model results in worse scores across all three architectural similarity metrics (worsening by 0.36 pp to 13.95 pp) and the encapsulation metrics (worsening by 2.08 pp). 
This outcome indicates that the LM, which can encode rich semantic information, enhances the recovery of Java modules.
\end{simplcolorbox}
\vspace{-1ex}

Finding 8 finding strongly support our insight that topic models leveraging LMs encode essential module recovery semantics, as topic models without LMs actually worsen \llm's performance. 
Overall, our findings for this RQ suggest that the key components of \llm and selecting just fully-qualified class names as its input type contribute significantly to its effectiveness.

\subsection{RQ5: Impact of Input Granularity} \label{input-granularity}

Unlike existing approaches, \llm purely relies on fully-qualified class names. To evaluate input representation impact, we experimented with varying granularity levels, from information-rich \textbf{code} (comprehensive source code) to minimal \textbf{package name} (package-level identifiers only).
In this setting, code includes the complete lexical content of each class---identifiers, comments, semantic variable names, and method invocations---while preserving their original frequencies to reflect richer semantic and behavioral information.

\subsubsection{Result Analysis}

\begin{table}[hbpt]
\centering
\scriptsize
\vspace{-1ex}
\caption{Impact of Input Granularity on Recovery Effectiveness: $\uparrow$ indicates a higher score is better, while $\downarrow$ indicates a lower score is better. The best score is \textbf{bolded}. 
- and + indicate value reduction or increase, respectively, for a metric.}
\vspace{-1ex}
\begin{tabular}{r|r|r|r}
\toprule
\hline
\rowcolor[RGB]{198,208,230}
\textbf{Metric}   & \textbf{\llm} & \textbf{Code} & \textbf{Package Name} \\
\hline
\textbf{a2a}       & \textbf{85.78\%} & -5.15 pp  & -1.67 pp       \\
\rowcolor[RGB]{224,230,242} \textbf{\cscore}   & \textbf{56.27\%} & -29.05 pp & -7.55 pp       \\
\textbf{\hscore}   & \textbf{77.44\%} & -39.86 pp & -1.97 pp       \\
\rowcolor[RGB]{224,230,242} \textbf{MQ}        & \textbf{15.61\%} & -7.60 pp  & -2.04 pp      
\\ \bottomrule                   
\end{tabular}
\label{table:input-study}
\vspace{-3ex}
\end{table}

\autoref{table:input-study} presents the experimental results. The performance metrics exhibit a consistent pattern: as input information increases (by including complete source code) or decreases (to only package names), both similarity and encapsulation metrics show varying degradation.

When using complete source code as input (including all identifiers and comments from classes and packages), we observe significant degradation in performance. The three metrics measuring architectural similarity with developer-created architectures decrease by 5.15--39.86 pp, while encapsulation metrics decline by 7.60 pp.
Similarly, reducing input to only package names leads to performance degradation across all four metrics, ranging from 1.67 to 7.55 pp.

The module-level architectural similarity metrics (\hscore{} and \cscore{}) show particular sensitivity to code-level input, likely due to the introduction of extraneous information. This suggests that achieving accurate architectural recovery at finer granularities is more susceptible to noise from excessive code-level details compared to system-level recovery.

\vspace{-1ex}
\begin{simplcolorbox}
Both increased (code) and decreased (package names) input granularity degraded \llm's recovery performance. The inclusion of complete code information yielded greater degradation, likely because architectural intent is less prominent in general code elements compared to fully-qualified class names. Package-name-only input produced modest degradation across all metrics, consistent with its semantic proximity to fully-qualified class names.
\end{simplcolorbox}
\vspace{-1ex}

The previous finding confirms that fully-qualified class names as input are essential for \llm to produce high-quality modules and contribute greatly to its effectiveness.

\vspace{-0.5ex}
\section{Threats to Validity} \label{sec:thread-to-validity}
\vspace{-0.5ex}
 
\textbf{\textit{External Threats.}} 
One threat to the external validity concerns the generalizability of the selected \jpms projects. To mitigate this threat, we employed a systematic project selection strategy, focusing on popular projects from GitHub that exhibit diversity across multiple dimensions: functionality domains, project sizes, and module counts. This selection approach helps ensure our subjects are representative of real-world \jpms applications.

\textbf{\textit{Internal Threats.}}
An internal threat to validity stems from our dependency tools. To mitigate the threats, we used Classycle for static dependency extraction, a well-established tool used in recent work~\cite{schmitt2020arcade,garcia2021constructing,ghorbani2019detection}.

To mitigate the issue brought by the embedding model,  we selected the open-source CodeBERT-java embedding model based on its popularity and specialized design for Java code representation.
Moreover, it is broadly applicable across Java projects rather than tailored to a specific domain, and it minimizes the risk of data leakage by being trained on publicly available datasets, and it operates as an encoder-only model to ensure stable, deterministic embeddings suitable for clustering-based recovery.
Our pilot study demonstrates that CodeBERT-java outperforms state-of-the-art encoder models (i.e., BERT~\cite{devlin2018bert}, CodeT5+~\cite{wang2023codet5}, OpenAI Embedding~\cite{openaiOpenAIPlatform}) on the most widely used architecture similarity score (a2a) by 1.56\% to 2.35\% when integrated with \llm, indicating that our methodology is not overly dependent on the chosen model.
The choice of using CodeBERT-java also ensures experimental reproducibility compared to closed-source alternatives (e.g., OpenAI).
We employed default parameter values for the general reliability of those dependencies.

Another internal threat arises from evaluating projects possibly used to train the model we use.
Upon examining the CodeBERT for Java training dataset~\cite{neulabcodebertjava}, we identified only a single module description file (i.e., \texttt{module-info.java}) from Cache2k, with no other modules from projects in our study present. Module description files typically contain only interfaces, excluding module internal classes.
Furthermore, the tasks for which that model was pre-trained~\cite{feng2020codebert, zhang2019bertscore} are unrelated to architecture recovery, minimizing the risk of data leakage impacting the evaluation.

Finally, a threat to validity arises from using modularized projects as evaluation subjects due to the challenge of finding identical project versions with only modularization differences. We mitigated this challenge by examining projects like JUnit5 and OpenJDK JDK. We found that over 90\% of source files in modularized versions have exact class name matches with their non-modularized counterparts. In other words, there are very less major structural refactoring activities during the modularization process, which align with prior studies~\cite{kim2012field,kim2014empirical}. Moreover, our successful module recovery using only \textit{class names}, without module-related information, further mitigates this threat.

\vspace{-0.5ex}
\section{Related Work} \label{sec:related-work}
\vspace{-0.5ex}

\textbf{JPMS Studies}. 
Most recently, A recent empirical study~\cite{bse} on JPMS reveals widespread instances of breaking strong encapsulation, where developers break module boundaries to access internal elements, leading to architectural inconsistency and migration challenges. By analyzing 4,079 GitHub issues, the study highlights the tension between enforced encapsulation and developers’ practical needs.
Despite the introduction of \jpms, many Java applications remain purely object-oriented (OO), as converting them to well-structured component-based (CB) systems presents significant challenges. OO2CB~\cite{hammad2022tool} enables conversion to least-privilege modularized systems using techniques like \llm. Moreover, recent tools address various JPMS challenges: Darcy~\cite{ghorbani2019detection,ghorbani2024darcy} addresses excessive module dependencies through automated detection and repair, ModGuard~\cite{dann2019modguard} enhances module security by identifying unauthorized access patterns, and Acadia~\cite{ghorbani2024bringing} implements runtime architecture-based adaptation without code modifications. For architectural evolution, Mondal et al.~\cite{mondal2021semantic} developed semantic slicing techniques to analyze structural changes in \jpms-based systems. However, prior research has not explored approaches for recovering Java modules from monolithic architectures. 

\textbf{Human Studies of Architecture Modularization}. Recent work has studied architectural challenges (e.g., architecture recovery) through interviews and surveys. 
Wan et al.~\cite{wan2023software} interviewed 32 software practitioners and identified key challenges in architectural modularization, encapsulation, coupling, and cohesion---areas that architecture recovery techniques like \llm aim to address.
Similarly, Tian et al.\cite{tian2022relationships} surveyed 87 respondents and interviewed 8 practitioners, finding that architecture recovery and tracing architectural constructs to code represent major challenges in practice.

\textbf{Non-JPMS Architecture Recovery}. Many studies have focused on architecture recovery in the past \cite{sar_2025,11203255,SemRef,ssar,ducasse2009software,koschke2009archrecon}. Recent LLM-based approaches build on this foundation by incorporating large language models (LLMs) to improve recovery quality. SSAR~\cite{ssar} enhances dependency-driven clustering using weighted graphs, while SemArc~\cite{11203255} combines code semantics, architectural-pattern semantics, and both explicit and implicit dependencies to more accurately group source files into meaningful architectural modules. SemRef~\cite{SemRef} further leverages LLMs to post-process recovered architectures. In contrast, \llm addresses a different problem: it extracts Java modules from monolithic systems without dependencies, source code, or prior recovery results, relying solely on fully-qualified class names and LM-based topic modeling with undersized-module repair. Because publicly available implementations of these three tools are not yet provided, we were unable to include them in our evaluation.

Amalfitano et al.~\cite{ECSALLM} evaluate state-of-the-art LLMs for software architecture recovery and finds that, while they can identify high-level structural and stylistic elements, they struggle with deeper abstractions such as class relationships and fine-grained design patterns.
The work done by Rukmono et al.~\cite{rukmono2024deductive} is fundamentally and conceptually distinct from our work: it envisions a top-down methodology, whereas our approach is bottom-up (i.e., inductive). Also, they do not include JPMS modules in their research plan; thus, the objective of our work is different from theirs. 
CAESAR~\cite{ibrahim2023context} is a context-aware software architecture recovery technique for C/C++. 
Similarly, SADE~\cite{papachristou2019software} can also recover the Linux kernel effectively with both semantic and structural information. However, CAESAR and SADE are not designed to recover Java-based software systems or are not publicly available.
Sadat et al. leverage a multi-objective optimization~\cite{sadat2019multi} to balance non-structural and structural features or cohesion and coupling~\cite{harman2010software}.
Boerstra et al.~\cite{boerstra2022stronger} presented a consensus-based approach for architectural recovery and the extraction of microservices. Their empirical findings showcased its superior performance over Bunch and Spectral clustering algorithms. 
However, the performance of the union- or consensus-group-based method was not better than using just static analysis~\cite{boerstra2022stronger}, so we excluded it.

Various studies of software architecture have been conducted using recovery techniques. 
Some frameworks, such as SAIN~\cite{garcia2021constructing} and ARCADE~\cite{schmitt2020arcade}, support a range of software recovery techniques and are designed to support empirical studies of software maintenance using recovered architectures \cite{le2015empirical, behnamghader2017large}. 
Lutellier et al.~\cite{lutellier2015comparing, lutellier2017measuring} conducted a study to measure the impact of code dependencies (e.g., dynamic and transitive dependencies) on the effectiveness of software architecture recovery techniques and showed accurate symbol dependencies improve the recovered architectures' quality.
Garcia et al. leveraged architecture recovery to build prediction models of architectural decay~\cite{garcia2021forecasting}. However, these studies have not addressed the specific context of \jpms, revealing a significant gap in the current literature by introducing a new, lightweight, and dependency-free recovery paradigm tailored to JPMS modules.

\vspace{-0.5ex}
\section{Conclusion and Future Work} \label{sec:conclusion}
\vspace{-0.5ex}
\jpms has been introduced to address encapsulation, modularization, and security issues that arise due to architectural decay. While existing architecture recovery tools have proven inadequate for Java module extraction, no prior research has investigated the systematic extraction of \jms from monolithic architectures.
To bridge this gap, we introduce \llm, a novel Java module extraction technique. Our empirical evaluation demonstrates \llm's superior performance compared to state-of-the-art architecture recovery techniques (i.e., ACDC, ARC SARIF, and FCA). When evaluated against the ground truth (i.e., \das), \llm achieves an improvement of 11.46 pp in a2a scores and exhibits enhanced encapsulation properties as measured by MQ scores. Moreover, \llm's computational efficiency surpasses existing approaches, executing 3.99 to 10.50 times faster than prior techniques.

While \jpms emphasizes both compile-time and run-time architecture, \llm primarily focuses on static architectural properties. Future work could incorporate dynamic information collected from run-time behavior to enhance the recovered software architecture.
Additionally, our approach recovers Java modules from monolithic architectures without defining module interfaces. Future work could address interface design, which requires understanding how external modules and client applications interact with specific Java modules across the broader ecosystem to properly balance security and encapsulation properties.

\section{Acknowledgments}
We sincerely appreciate Prof. André van der Hoek and Prof. Sam Malek for their constructive feedback and support.

\scriptsize
\bibliographystyle{IEEEtran}
\bibliography{references}

\end{document}